\documentclass[11pt]{article}
\usepackage{amssymb,amsfonts,amsmath, psfrag,eepic,graphicx,float}

\makeatletter \@addtoreset{equation}{section}
\@addtoreset{theo}{section} \makeatother

 \makeatletter
\newfont{\footsc}{cmcsc10 at 8truept}
\newfont{\footbf}{cmbx10 at 8truept}
\newfont{\footrm}{cmr10 at 10truept}

\title{\bf Explicit quasi-periodic wave solutions and  asymptotic
analysis to  the   supersymmetric Ito's  equation}

\author{Engui Fan$^{a}$\footnote{  Electronic mail:
faneg@fudan.edu.cn.},  \ \  Y. C. Hon$^{b}$,
\\
\small  a. School of Mathematics Sciences, Fudan University, Shanghai 200433, PR China\\
\small  b. Department of Mathematics,  City University of Hong Kong,
Hong Kong, PR China\\}
\begin{document}
\maketitle
 \baselineskip=18pt
\noindent {\large \bf Abstract:}  Based on a Riemann theta function
and the super-Hirota bilinear form, we propose  a key formula for
explicitly constructing quasi-periodic  wave solutions of the
supersymmetric Ito's equation in superspace
$\mathbb{C}_{\Lambda}^{2,1}$. Once a nonlinear equation is written
in bilinear forms, then the quasi-periodic wave solutions can be
directly obtained from our formula. The relations between the
periodic wave solutions and the well-known soliton solutions are
rigorously established.  It is shown that the quasi-periodic wave
solutions tends to the  soliton solutions under small amplitude
limits.

\section{Introduction}
The Ito's equation takes the form
$$u_{tt}+6(u_xu_t)_x+u_{xxxt}=0,\eqno(1.1)$$
which was first proposed by Ito,  and its bilinear B\"{a}cklund
transformation, Lax representation and multi-soliton solutions were
 obtained \cite{Ito} . The other integrable properties of this equation
 such as the nonlinear superposition formula, Kac-Moody algebra,
bi-Hamiltonian structure have been further found
\cite{Hu1}-\cite{Liu1} . Recently, Liu, Hu and Liu proposed  the
following supersymmetric Its's equation \cite{Liu2}
$$\begin{aligned}
&\mathfrak{D}_tF_t+6(F_x(\mathfrak{D}_tF))_x+\mathfrak{D}_tF_{xxx}=0,
\end{aligned}\eqno(1.2)$$
and obtained its one-, two- and three-soliton solutions,  where
$F=F(x,t,\theta)$ is fermionic superfield depending on usual even
independent variable $x$, $t$ and odd Grassman variable $\theta$.
The differential operator
$\mathfrak{D}_t=\partial_{\theta}+\theta\partial_t$
 is the super derivative.

  The bilinear derivative method developed by Hirota is a powerful
approach for constructing exact solution of nonlinear
equations\cite{Hirota1}--\cite{Sa} . Based on the Hirota bilinear
form and the Riemann theta functions,  Nakamura presented an
approach to directly construct a kind of quasi-periodic solutions of
nonlinear equation \cite{Na1,Na2} , where  the periodic  wave
solutions of the KdV equation and the Boussinesq equation were
obtained. This method not only conveniently obtains periodic
solutions of a nonlinear equation, but also directly gives the
explicit relations among frequencies, wave-numbers, phase shifts and
amplitudes of the wave. Recently, this method is further developed
to investigate the discrete Toda lattice, (2+1)-dimensional
Kadomtsev-Petviashvili equation  and Bogoyavlenskii's   breaking
soliton equation\cite{Dai}-\cite{Fan2} .

 Our present paper will  considerably improve the key steps
  of the above method so as to   make the method  much more lucid and straightforward for applying
   a class of  nonlinear supersymmetric equations.   First,
 the above method will be generalized into the supersymmetric
context.  The quasi-periodic solutions of supersymmetric  equations
still seem  not investigated to our acknowledge.   Second, we a
formula that the Riemannn theta functions satisfy a super-Hirota
bilinear equation. This formula actually provides us an  uniform
method which can be used to construct quasi-periodic wave solutions
of nonlinear differential, difference and supersymmetric equations.
Once a nonlinear equation is written in bilinear forms, then the
  quasi-periodic wave solutions of the nonlinear equation can be obtained directly
  by using the formula. As illustrative example, we shall
  construct quasi-periodic wave   solutions to the supersymmetric Ito's equation (1.2).
    Moreover, we also establish the relations between our quasi-periodic wave
solutions  and  the   soliton solutions that were obtained by Liu
and Hu \cite{Liu1} .

  The organization of this paper is as follows. In section 2, we briefly introduce a
 super-Hirota bilinear that will be suitable  for
constructing quasi-periodic  solutions of the equation (1.2). And
then introduce a general  Riemann theta function  and  provide a key
formula for constructing periodic wave solutions.
  In section 3,  as application of our
formula, we construct one-periodic wave solutions to the equation
(1.2). We further present  a simple and effective  limiting
procedure to analyze  asymptotic behavior of the one-periodic wave
solutions.  It is rigorously shown that the quasi-periodic wave
solutions tends to the known  soliton solutions obtained by Liu and
Hu  under  ``small amplitude" limits.  At last, we briefly discuss
the conditions on the construction of multi-periodic wave solutions
of the equation (1.2) in section 4.

\section{  The superspace, Hirota bilinear form  and  the Riemann theta functions}

To fix the notations  and make our presentation self-contained,  we
briefly recall  some properties about superanalysis  and
super-Hirota bilinear operators.  The details about superanalysis
refer, for instance, to Vladimirov's work \cite{Vlad1, Vlad2}.

A linear space $\Lambda$ is called $Z_2$-graded if it represented as
a direct sum of two subspaces
$$\Lambda=\Lambda_0\oplus\Lambda,$$
where elements of the spaces $\Lambda_0$ and $\Lambda_1$ are
homogeneous. We assume that  $\Lambda_0$ is a subspace consisting of
even elements  and  $\Lambda_1$ is a subspace consisting of odd
elements. For the element $f\in\Lambda$ we denote by $f_0$ and $f_1$
its even and odd components. A parity function is introduced on the
$\Lambda$, namely,
$$|f|=\left\{\begin{matrix}0, \ \ {\rm if} \ \ f\in \Lambda_0,
\\ \ 1, \ \ {\rm if }\ \ f\in \Lambda_1.\end{matrix}\right.
$$

We introduce an annihilator of the set of odd elements by setting
$$^{\perp}\Lambda_1=\{\lambda\in\Lambda: \lambda\Lambda_1=0\}.$$

A superalgebra  is a $Z_2$-graded space
$\Lambda=\Lambda_0\oplus\Lambda$ in which, besides usual operations
of addition and multiplication by numbers, a product of elements is
defined with the usual distribution law:
$$a(\alpha b+\beta c)=\alpha ab+\beta ac, \ \ (\alpha b+\beta
c)a=\alpha ba+\beta ca,$$ where $a, b, c\in\Lambda$ and $\alpha,
\beta\in \mathbb{C}.$  Moreover,  a structure on $\Lambda$ is
introduced of an associative algebra with a unite $e$ and even
multiplication i.e., the product of two even and two odd elements is
an even element and the product of an even element by an odd one is
an odd element: $|ab|=|a|+|b|$ mod (2).

 A commutative superalgebra with unit $e=1$ is
called a finite-dimensional  Grassmann algebra if it contains a
system of anticommuting generators $\sigma_j, j=1, \cdots, n$ with
the property: $\sigma_j\sigma_k+\sigma_k\sigma_j=0, \ j,
k=1,2,\cdots, n$, in particular, $\sigma_j^2=0$. The Grassmann
algebra will be denote by $G_n=G_n(\sigma_1,\cdots, \sigma_n)$.

The monomials $\{e_0, e_i=\sigma_{j_1}\cdots\sigma_{j_n}\}$,
$j=(j_1<\cdots<j_n)$ form a basis in the Grassmann algebra $G_n$,
$\dim G_n=2^n$.  Then it follows that any element of $G_n$ is a
linear combination of monomials $\sigma_{j_1}\cdots\sigma_{j_k}, \
j_1<\cdots<j_k$, that is,
$$f=f_0+\sum_{k\geq 0}\sum_{j_1<\cdots<j_k}f_{j_1\cdots j_k}\sigma_{j_1}\cdots\sigma_{j_k},$$
where the coefficients $f_{j_1\cdots j_k}\in \mathbb{C}$.

{\bf Definition 1.}  Let $\Lambda=\Lambda_0\oplus\Lambda$ be a
commutative Banach superalgebra, then the Banach space
$$\mathbb{C}_{\Lambda}^{m,n}=\Lambda_0^m\times\Lambda_1^n$$
is called a superspace of dimension $(m,n)$ over $\Lambda$.  In
particular, if $\Lambda_0=\mathbb{C}$ and $\Lambda_1=0$, then
$\mathbb{C}_{\Lambda}^{m,n}=\mathbb{C}^m.$

 A function $f(\boldsymbol{x}):
\mathbb{C}_{\Lambda}^{m,n}\rightarrow \Lambda$ is said to be
superdifferentiable at the point $x\in \mathbb{C}_{\Lambda}^{m,n}$,
if there exist elements $F_j(\boldsymbol{x})$ in $\Lambda, \ j=1,
\cdots, m+n$, such that
$$f(\boldsymbol{x}+\boldsymbol{h})
=f(\boldsymbol{x})+\sum_{j=1}^{m+n}\langle
F_j(\boldsymbol{x}),h_j\rangle+o(\boldsymbol{x},\boldsymbol{h}),$$
where $\boldsymbol{x}=(x_1, \cdots, x_m, x_{m+1}, \cdots, x_{n})$
with components $ x_j, j=1,\cdots, m$ being even variable and
$x_{m+j}=\theta_j, j=1,\cdots, n$  being Grassmann odd ones. The
vector $\boldsymbol{h}=(h_1, \cdots, h_m$, $h_{m+1}, \cdots,
h_{m+n})$ with $ (h_1, \cdots, h_m)\in\Lambda_0^m$ and $(h_{m+1},
\cdots, h_{m+n})\in\Lambda_1^n$. Moreover,
$$\lim_{\parallel \boldsymbol{h}\parallel\rightarrow 0}\frac{\parallel o(\boldsymbol{x},\boldsymbol{h})
\parallel}{\parallel \boldsymbol{h}\parallel}
\longrightarrow 0.$$ The $F_j(\boldsymbol{x})$ are called the super
partial derivative of $f$ with respect to $x_j$ at the point
$\boldsymbol{x}$ and are denoted, respectively, by
$$\frac{\partial f(\boldsymbol{x})}{\partial x_j}=F_j(\boldsymbol{x}),\ j=1,\cdots, m+n.$$
The derivatives $\frac{\partial f(\boldsymbol{x})}{\partial x_j}$
with respect to even variables  $x_j, \ j=1,2,\cdots n$ are uniquely
defined. While the derivatives $\frac{\partial
f(\boldsymbol{x})}{\partial \theta_j}$ to odd variables
$\theta_j=x_{j+n}, \ j=1,2,\cdots m$ are not uniquely defined, but
with an accuracy to within an addition constant
$c\sigma_1\cdots\sigma_n, c\in \mathbb{C}$ from an annihilator
$^\perp G_n$ of finite-dimensional Grassmann algebra $G_n$.

The super derivative also satisfies Leibniz formula
$$\frac{\partial (f(\boldsymbol{x})g(\boldsymbol{x}))}{\partial x_j}=\frac{\partial f(\boldsymbol{x})}
{\partial x_j}g(\boldsymbol{x})
+(-1)^{|x_j||f|}f(\boldsymbol{x})\frac{\partial
g(\boldsymbol{x})}{\partial x_j}, \ j=1, \cdots, m+n.\eqno(2.1)$$

 Denote by  $\mathcal{P}(\Lambda_1^n,
\Lambda)$ the set of polynomials defined on $\Lambda_1^n$ with value
in $\Lambda$. We say that  a super integral is a map $I:
\mathcal{P}(\Lambda_1^n, \Lambda)\rightarrow \Lambda$ satisfying the
following condition is an super integral about Grassmann variable

(1) A linearity: $I(\mu f+\nu g)=\mu I(f)+\nu I(g), \ \mu,
\nu\in\Lambda, \ f, g\in\mathcal{P}(\Lambda_1^n, \Lambda); $

(2) translation invariance: $I(f_{\xi})=I(f)$, where
$f_{\xi}=f(\boldsymbol{\theta}+\boldsymbol{\xi})$ for all
$\boldsymbol{\xi}\in\Lambda_1^n$, $f\in\mathcal{P}(\Lambda_1^n,
\Lambda).$

We denote $I(\theta^\varepsilon)=I_\varepsilon$, where $\varepsilon$
belongs to the set of multiindices
$N_n=\{\boldsymbol{\epsilon}=(\varepsilon_1, \cdots, \varepsilon_n),
\varepsilon_j=0,1,
\boldsymbol{\theta}^\varepsilon=\theta_1^{\varepsilon_1}\cdots\theta_n^{\varepsilon_n}\not\equiv
0\}$. In the case when $I_\varepsilon=0, \varepsilon\in N_n,
|\varepsilon|\leq n=n-1$, such kind of integral has the form
$$I(f)=J(f)I(1,\cdots,1),$$
where
$$J(f)=\frac{\partial^nf(0)}{\partial\theta_1\cdots\partial\theta_n}.$$
Since the derivative is defined with an accurcy to with an additive
constant form the annihilator $^\perp L_n$,
$L_n=\{\theta_1\cdots\theta_n, \boldsymbol{\theta}\in
\Lambda_1^n\}$, it follows that $J:\mathcal{P}\rightarrow
\Lambda/^\perp L_n$ is single-valued mapping. This mapping also
satisfies the conditions 1 and 2, and therefore we shall call it an
integral and denote
$$J(f)=\int f(\boldsymbol{\theta})d\boldsymbol{\theta}=\int\theta_1\cdots\theta_n d\theta_1\cdots d\theta_n,$$
which has properties:
$$\begin{aligned}
&\int\theta_1\cdots\theta_n d\theta_1\cdots d\theta_n=1,\\
&\int\frac{\partial f}{\partial\theta_j}d\theta_1\cdots
d\theta_n=0,\ j=1, \cdots, n.\\
&\int f(\boldsymbol{\theta})\frac{\partial
g(\boldsymbol{\theta})}{\partial\theta_j}d\boldsymbol{\theta}=
(-1)^{1+|g|}\int \frac{\partial
f(\boldsymbol{\theta})}{\partial\theta_j}g(\boldsymbol{\theta})d\boldsymbol{\theta}.
\end{aligned}\eqno(2.2)$$

In this paper, we consider functions with two ordinary even
variables $x, t$ and a Grassmann odd variable $\theta$. The
associated  space
$\mathbb{C}_{\Lambda}^{2,1}=\Lambda_0^2\times\Lambda_1$ (we may take
$\Lambda_0=\mathbb{R}$ or $\mathbb{C}$) is a superspace over
Grassmann algebra $G_1(\sigma)=G_{1,0}\oplus G_{1,1}$, whose
elements have the form
 $$f=f_0+f_1\sigma.$$
where $e=1$ is a unit, $\sigma$ is anticommuting generator.
 The monomials $\{1, \sigma\}$ form a basis of the
$G_1(\sigma)$, dim$G_1(\sigma)=2$.   Under traveling wave frame in
space $\mathbb{C}_{\Lambda}^{2,1}$, the
 phase variable  should have  the form
$$\xi=\alpha x+\omega t+\theta\sigma. $$

   Now we consider the bilinear form of the equation (1.2). By the dependent variable transformation
$$F=\partial_x\ln f(x,t,\theta), \eqno(1.3)$$
where $f(x,t,\theta): \mathbb{C}_{\Lambda}^{2,1}\rightarrow
\mathbb{C}_{\Lambda}^{1,0}$ is a superdifferential function, the
equation (1.2) is then transformed into a bilinear form
$$(S_tD_t+S_tD_x^3)f(x,t,\theta)\cdot f(x,t,\theta)=0.\eqno(1.4)$$
where the Hirota bilinear differential operators $D_x$  and $D_t$
are defined by
\begin{eqnarray*}
&&D_x^mD_t^n f(x,t,\theta)\cdot g(x, t,
\theta)=(\partial_x-\partial_{x'})^m(\partial_t-\partial_{t'})^n
f(x, t, \theta) g(x', t', \theta')|_{x'=x, t'=t, \theta'=\theta}.
\end{eqnarray*}
 The super-Hirota bilinear operator is defined as
\cite{Carstea}
$$ S_t^N f(x,t,\theta)\cdot
g(x,t,\theta)=\sum_{j=0}^N(-1)^{j|f|+\frac{1}{2}j(j+1)}\left[\begin{matrix} N\\
j\end{matrix}\right]\mathfrak{D}_t^{N-j}f(x,t,\theta)\mathfrak{D}_t^jg(x,t,\theta),$$
 where   the super
binomial coefficients are defined by
$$\left[\begin{matrix} N\\
j\end{matrix}\right]=\left\{\begin{matrix}\left(\begin{matrix}
[N/2]\cr [j/2]\end{matrix}\right), {\rm if} \ \ (N, j)\not=(0,1)\ \
{\rm mod}\ \ 2,\\  0, \ \ {\rm otherwise}.\end{matrix}\right.$$
$[k]$ is the integer part of the real number $k$ ($[k]\leq k\leq
[k]+1$).

Following the Hirota bilinear theory,  It is easy to find that the
equation (1.2) admits one-soliton solution (also called
one-supersoliton solution)
$$F_1=\partial_x\ln(1+e^{\eta}),\eqno(1.5)$$
with  phase variable $\eta= k x-k^3 t+\theta \zeta+\gamma$ and $k,
\gamma\in \Lambda_0$, $\zeta\in \Lambda_1$.

To apply the Hirota bilinear method for constructing periodic wave
solutions of the equation (1.2),   we hope to  add  two odd
variables $F_0$, $c$   and consider a more general  form than the
bilinear equation (1.4)

$$F=\partial_{\theta}^{-1}F_0+\partial_x \ln f(x, t, \theta),\eqno(2.1)$$
where  $F_0=F_0(\theta): \mathbb{C}_{\Lambda}^{2,1}\rightarrow
\mathbb{C}_{\Lambda}^{0,1}$ is an odd special  solution of the
equation (1.2). Substituting (2.1) into (1.2) and integrating with
respect to $x$, we then get the following bilinear form
$$\begin{aligned}
&G(S_t, D_x,D_t)f\cdot f=(S_tD_t+S_tD_x^3+3F_0D_x^2+c) f\cdot f=0,
\end{aligned}\eqno(2.2)$$ where
 $c=c(\theta,t): \mathbb{C}_{\Lambda}^{2,1}\rightarrow
\mathbb{C}_{\Lambda}^{0,1}$ is an odd integration constant.   For
the bilinear equation (2.2), we are interested in its multi-periodic
solutions in terms of the Riemann theta functions.

In the following,  we introduce a super one-dimensional Riemann
theta function on super space $\mathbb{C}_{\Lambda}^{2,1}$ and
discuss its quasi-periodicity, which plays a central role in this
paper.  The Riemann theta function reads
$$
\vartheta(\mathbf{\xi},  \varepsilon,s|\tau)=\sum_{n\in
\mathbb{Z}}\exp[2\pi i(\xi+\varepsilon)(n+s)-\pi \tau
({n}+{s})^2].\eqno(2.3)$$ Here the integer value $n\in \mathbb{Z}$,
$s, \varepsilon\in \mathcal{C}$, and complex phase variables
$\xi=\alpha x+\omega t+\theta\sigma+\delta$ is dependent of even
variable $x, t$ and odd $\theta$; The $\tau>0$ is called the period
matrix of the Riemann theta function. It is obvious that the Riemann
theta function (2.3)  converges absolutely and superdifferentiable
on superspace $\mathbb{C}_{\Lambda}^{2,1}$.
 For
the simplicity, in the case when $s=\omega=0$, we denote
$$\vartheta(\mathbf{\xi},\tau)=\vartheta(\mathbf{\xi},0, 0|\tau).$$

{\bf Definition 2.}  A function  $f(\xi):
\mathbb{C}_{\Lambda}^{2,1}\rightarrow \mathbb{C}_{\Lambda}^{1,0}$ is
said to be quasi-periodic in $\xi=\alpha x+\omega
t+\theta\sigma+\delta$ with fundamental periods  $T$, if there exist
certain constants $a, b\in \Lambda_0$,  such that
$$ f(\xi+T)=f(\xi)+a\xi+b. $$
An example of this is the ordinary Weierstrass zeta function, where
$$\zeta(\xi+\omega)=\zeta(\xi)+\eta,$$
for a fixed constant $\eta$ when $\omega$ is a period of the
corresponding Weierstrass elliptic $\wp$  function.

 {\bf Proposition 2.}  \cite{Far}  The Riemann theta function $\vartheta(\xi,\tau)$
 defined above  has the periodic properties
$$\begin{aligned}
&\vartheta(\xi+1+i\tau,\tau)=\exp(-2\pi i\xi+\pi
\tau)\vartheta(\xi,\tau).
\end{aligned}\eqno(2.4)$$

Now we turn to see the periodicity of the solution (2.4), we take
 $f(x,t,\theta)$  in the bilinear equation (2.2) as
 $$f(x,t,\theta)=\vartheta(\xi, \tau),$$
where  phase variable $\xi=\alpha x+\omega t+\theta\sigma +\delta$.
By using (2.4), it is easy to see that
$$\begin{aligned}
&\frac{\vartheta'_{\xi}(\xi+i\tau,\tau)}{\vartheta(\xi+i\tau,\tau)}
=-2\pi i+\frac{\vartheta'_{\xi}(\xi,\tau)}{\vartheta(\xi,\tau)},
\end{aligned}$$
that is,
$$\begin{aligned}
&\partial_{\xi}\ln\vartheta(\xi+i\tau,\tau)=-2\pi
i+\partial_{\xi}\ln \vartheta(\xi,\tau).
\end{aligned}\eqno(2.5)$$
According to the differential relation, we have
$$F(x,t,\theta)=F(\xi)=\partial_{\theta}^{-1}F_0+\alpha
\partial_{\xi}\ln\vartheta(\xi,\tau).\eqno(2.6)$$
The equations (2.5) and (2.6) demonstrate that
$$F(\xi+1+i\tau)=\partial_{\theta}^{-1}F_0+\alpha
\partial_{\xi}\ln\vartheta(\xi+1+i\tau,\tau)=-2\pi i\alpha+F(\xi).$$
Therefore  the solution  $F(\xi)$ is a quasi-periodic function with
two fundamental periods $1$ and $i\tau$.

In following, we establish uniform formula on the Riemannn theta
function, which will play a key role in the construction of the
periodic wave solutions.

{\bf Proposition 2.} \cite{Carstea}  Suppose that $f(x,t,\theta),
g(x,t,\theta)$ are super differentiable on space
$\mathbb{C}_{\Lambda}^{2,1}$. Then the Hirota bilinear operators
$D_x, D_t$ and super-Hirota bilinear operator $S_x$ have properties
$$\begin{aligned}
&S_x^{2N}f\cdot
  g =D_x^N f\cdot g,\\
&D_x^mD_t^n   e^{\xi_1}\cdot
  e^{\xi_2}=(\alpha_1-\alpha_2)^m(\omega_1-\omega_2)^n
  e^{\xi_1+\xi_2},\\
 &S_x e^{\xi_1}\cdot
  e^{\xi_2}=[\sigma_1-\sigma_2+\theta(\alpha_1-\alpha_2)]
  e^{\xi_1+\xi_2},
\end{aligned}\eqno(2.7)$$
where  $\xi_j=\alpha_jx+\omega_jt+\theta\sigma_j+\delta_j, j=1,2$.
More generally, we have
$$\begin{aligned}
&F(S_x, D_x, D_t)e^{\xi_1}\cdot
  e^{\xi_2} =F(\sigma_1-\sigma_2+\theta(\alpha_1-\alpha_2),\alpha_1-\alpha_2,\omega_1-\omega_2)
  e^{\xi_1+\xi_2},\end{aligned}\eqno(2.8)$$
where $G(S_t, D_x, D_t)$ is a polynomial about $S_t, D_x$ and $
D_t$. This properties  will be utilized  later to explore
  the quasi-periodic wave solutions of the equation (1.2).

{\bf Proposition 3.}  The Hirota bilinear operators  $D_x, D_t$ and
super-Hirota bilinear operator   $S_x$ have  properties when they
act on the Riemann theta functions
$$D_x \vartheta(\xi,\varepsilon', s'|\tau)\cdot
  \vartheta(\xi,\varepsilon, s|\tau)=\sum_{\mu=0,1}\partial_x\vartheta(2\xi,\varepsilon'-\varepsilon,
  (s'-s-\mu)/2|2\tau)|_{\xi=0}\vartheta(2\xi,\varepsilon'+\varepsilon,(s'+s+\mu)/2|2\tau),\eqno(2.9)$$
$$S_t  \vartheta(\xi,\varepsilon', s'|\tau)\cdot
  \vartheta(\xi,\varepsilon, s|\tau)=\sum_{\mu=0,1}\mathfrak{D}_t\vartheta(2\xi,\varepsilon'-\varepsilon,
  (s'-s-\mu)/2|2\tau)|_{\xi=0}\vartheta(2\xi,\varepsilon'+\varepsilon,(s'+s+\mu)/2|2\tau),
\eqno(2.10)$$ where $\sum_{\mu=0,1}$ indicates sum with respective
to  $\mu=0,1$.

 In general, for a polynomial operator $G(S_t, D_x,
D_t)$ about $S_t, D_x$ and $ D_t$, we have
$$\begin{aligned}
&G(S_t, D_x, D_t)\vartheta(\xi,\tau)\cdot
  \vartheta(\xi,\tau)=\sum_{\mu=0,1}C(\alpha,\omega, \sigma, \mu) \vartheta(2\xi,\mu/2|2\tau),\end{aligned}\eqno(2.11)$$
where
$$\begin{aligned}
&\xi=\alpha x+\omega t+\theta\sigma+\gamma.\\
  &C(\alpha,\omega, \sigma, \mu|\tau)= \sum_{n\in Z}
  G\left\{4\pi i(n-\mu/2)\alpha, 4\pi
  i(n-\mu/2)\omega,\right.\\
 & \left.  4\pi i(n-\mu/2)(\sigma+\theta\omega) \right\}\times \exp\left[-2\pi\tau(n-\mu/2)^2\right].
\end{aligned}\eqno(2.12)$$

{\it Proof.} By using (2.7), we have
\begin{eqnarray*}
&&\Gamma=S_t  \vartheta(\xi,\varepsilon',s'|\tau)\cdot
  \vartheta(\xi,  \varepsilon, s|\tau)\\
  &&=\sum_{m',m\in \mathbb{Z}}S_t\exp\{2\pi i(m'+s')(\xi+\varepsilon')-\pi(m'+s')^2\tau\}\cdot
  \exp\{2\pi i(m+s)(\xi+\varepsilon)-\pi(m+s)^2\tau\},\\
  &&=\sum_{m',m\in \mathbb{Z}}2\pi i(\sigma+\theta\omega)(m'-m+s'-s) \exp\left\{2\pi
  i(m'+m+s'+s)\xi-2\pi i[(m'+s')\varepsilon'+(m+s)\varepsilon]\right.\\
  &&\ \ \ \ \ \ \ \ \ \ \  \ \   \
  \left.-\pi\tau[(m'+s')^2+(m+s)^2]\right\}\\
  &&\stackrel{m=l'-m'}{=}\sum_{l',m'\in \mathbb{Z}}2\pi i(\sigma+\theta\omega)(2m'-l'+s'-s) \exp\left\{2\pi
  i(l'+s'+s)\xi-2\pi i[(m'+s')\varepsilon'\right.\\
&&\ \ \ \ \ \ \ \ \ \ \  \ \   \
  \left. +(l'-m'+s)\varepsilon]-\pi[(m'+s')^2+(l'-m'+s)^2]\tau\right\}\\
  &&\stackrel{l'=2l+\mu}{=}\sum_{\mu=0,1}\sum_{l,m'\in \mathbb{Z}}2\pi i(\sigma+\theta\omega)(2m'-2l+s'-s-\mu)\exp\{4\pi
  i\xi[l+(s'+s+\mu)/2]\\
  &&\ \ \ \ \   -2\pi i[(m'+s')\varepsilon'-(m'-2l-s-\mu)\varepsilon] -\pi[(m'+s')^2+(m'-2l-s-\mu)^2]\tau\}
  \end{eqnarray*}
Let $m'=n+l$, and using the relations
\begin{eqnarray*}
    &&n+l+s'=[n+(s'-s-\mu)/2]+[l+(s'+s+\mu)/2],\\
   &&n-l-s-\mu=[n+(s'-s-\mu)/2]-[l+(s'+s+\mu)/2],
\end{eqnarray*}
 we finally  obtain that
{\small\begin{eqnarray*}
    &&\Gamma=\sum_{\mu=0,1}\left[\sum_{n\in \mathbb{Z}}4\pi i(\sigma+\theta\omega)[n+(s'-s-\mu)/2]
    \exp\{-2\pi i[n+(s'-s-\mu)/2](\varepsilon'-\varepsilon)-2\pi\tau[n+(s'-s-\mu)/2]^2\}\right]\\
&& \ \ \ \ \ \ \ \ \times \left[\sum_{l\in \mathbb{Z}}\exp\{2\pi
i[l+(s'+s+\mu)/2](2\xi+\varepsilon'+\varepsilon)-2\pi\tau[l+(s'+s+\mu)/2]^2\right]\\
&&=\sum_{\mu=0,1}\mathfrak{D}_t\vartheta(2\xi,\varepsilon'-\varepsilon,
(s'-s-\mu)/2|2\tau)|_{\xi=0}
  \vartheta(2\xi,\varepsilon'+\varepsilon, (s'+s+\mu)/2|2\tau).
\end{eqnarray*}}

In a similar way, we can prove the formulae (2.9). As a special case
when $\varepsilon=s=0$ of the Riemann theta function (2.3), by using
(2.9) an  (2.10),  we can prove the formula (2.11). $\Box$

  From the formulae (2.11) and (2.12), it is seen that if  the following equations are
  satisfied
  $$C(\alpha,\omega, \sigma, \mu|\tau)=0,$$
  for  $\mu=0,1$, then
  $\vartheta(\xi,\tau)$ is a solution of the bilinear equation
  $$G(S_t, D_x, D_t)\vartheta(\xi,\tau)\cdot\vartheta(\xi,\tau)=0.$$
\section{ Quasi-periodic waves and  asymptotic properties}

 In this section, we  consider periodic wave solutions of the equation
 (1.2).  As a  simple case of  the theta function
 (2.3)  when $N=1, s=0$,  we take $f(x,t,\theta)$ as
$$f(x,t,\theta)=\vartheta(\xi,\tau)=\sum_{n\in \mathbb{Z}}\exp({2\pi in\xi-\pi n^2\tau}),\eqno(3.1)$$
where the phase variable $\xi=\alpha x+\omega
t+\theta\sigma+\delta$, and the parameter $\tau>0$.

To let the Riemann theta function (3.1) be a solution of the
bilinear equation (2.2),  according to the formula (2.11), the
following equations only need to be  satisfied
 $$\begin{aligned}
&\sum_{n\in \mathbb{Z}}\left[-16\pi^2
(n-\mu/2)^2(\sigma+\theta\omega)\omega+256\pi^4(n-\mu/2)^4(\sigma+\theta\omega)\alpha^3\right.\\
&\left. \ \ \ \ \ \ \ \ \
-48\pi^2(n-\mu/2)^2\alpha^2F_0+c\right]\exp[-2\pi
(n-\mu/2)^2\tau]=0,\ \mu=0, 1.
\end{aligned}\eqno(3.2)$$

We introduce the notations by
$$\begin{aligned}
&\lambda=e^{-\pi\tau/2 },\quad
\vartheta_1(\xi,\lambda)=\vartheta(2\mathbf{\xi},2\tau)=\sum_{n\in\mathbb{Z}}
\lambda^{4n^2}\exp(4i\pi n\xi),\\
&\vartheta_2(\xi,\lambda)=\vartheta(2\xi, 0,
-1/2,2\tau)=\sum_{n\in\mathbb{Z}} \lambda^{(2n-1)^2}\exp[2i\pi(2n-1)
\xi].\end{aligned}\eqno(3.3)$$

By  using formula (3.3),  the equation (3.2) can be written as  a
linear system
 $$\begin{aligned}
&\theta\vartheta_1''\omega^2+(\sigma\vartheta_1''+\alpha^3\theta\vartheta_1^{(4)})
\omega+\vartheta_1c+\sigma\alpha^3\vartheta_1^{(4)}+3\alpha^2F_0\vartheta_1''=0,\\
&\theta\vartheta_2''\omega^2+(\sigma\vartheta_2''+\alpha^3\theta\vartheta_2^{(4)})
\omega+\vartheta_2c+\sigma\alpha^3\vartheta_2^{(4)}+3\alpha^2F_0\vartheta_2''=0,
\end{aligned}\eqno(3.4)$$
where $\omega\in\Lambda_0$ is even and $c, F_0:
\mathbb{C}_{\Lambda}^{2,1}\rightarrow \mathbb{C}_{\Lambda}^{0,1}$
are odd.  In addition,  we have denoted   derivatives  of
$\vartheta_j(\xi,\lambda)$ at $\xi=0$  by simple notations
$$\vartheta_j^{(k)}=\vartheta_j^{(k)}(0,\lambda)=\frac{d^k\vartheta_j(\xi,\lambda)}{d\xi^k}|_{\xi=0}, \ \
j=1,2, k=0, 1,2,\cdots$$ Moreover, these functions are independent
of Grassmann variable $\theta$ and $\sigma$.

We show there existence real solutions to the system (3.4). Since
$c=c(\theta, t)$ and $F=F_0(\theta)$ are function of Grassmann
variable $\theta$,  we can expand them in the form
$$c=c_1+c_2\theta, \ \ F_0=f_1+f_2\theta,\eqno(3.5)$$
where $c_1, f_1\in\Lambda_1$ are odd and  $c_2, f_2\in\Lambda_0$ are
even. Substituting (3.5) into (3.4) leads to
 $$\begin{aligned}
&(\sigma\vartheta_1''\omega+\vartheta_1c_1+\sigma\alpha^3\vartheta_1^{(4)}+3\alpha^2\vartheta_1''f_1)
+\theta(3\alpha^2\vartheta_1''f_2+\vartheta_1c_2+\vartheta_1''\omega^2+\alpha^3\vartheta_1^{(4)}\omega)
=0,\\
&(\sigma\vartheta_2''\omega+\vartheta_2c_1+\sigma\alpha^3\vartheta_2^{(4)}+3\alpha^2\vartheta_2''f_1)
+\theta(3\alpha^2\vartheta_2''f_2+\vartheta_2c_2+\vartheta_2''\omega^2+\alpha^3\vartheta_2^{(4)}\omega)=0,
\end{aligned}\eqno(3.6)$$
where $\omega,\ c_1, \ c_2, f_1$ and $f_2$ are parameters to be
determined.

Since $\theta$ is a  Grassmann variable,  the  system (3.6) will be
satisfied provided that
 $$\begin{aligned}
&\sigma\vartheta_1''\omega+\vartheta_1c_1+\sigma\alpha^3\vartheta_1^{(4)}+3\alpha^2\vartheta_1''f_1
=0,\\
&\sigma\vartheta_2''\omega+\vartheta_2c_1+\sigma\alpha^3\vartheta_2^{(4)}+3\alpha^2\vartheta_2''f_1=0
\end{aligned}\eqno(3.7)$$
and
$$\begin{aligned}
&3\alpha^2\vartheta_1''f_2+\vartheta_1c_2+\vartheta_1''\omega^2+\alpha^3\vartheta_1^{(4)}\omega
=0,\\
&3\alpha^2\vartheta_2''f_2+\vartheta_2c_2+\vartheta_2''\omega^2+\alpha^3\vartheta_2^{(4)}\omega=0.
\end{aligned}\eqno(3.8)$$

In the systems (3.7) and (3.8), it is obvious that vectors
$(\vartheta_1, \vartheta_2)^T$ and $ (\vartheta_1'',
\vartheta_2'')^T$ are linear independent, and $(\vartheta_1^{(4)},
\vartheta_2^{(4)})^T\not=0$.  Therefore the system (3.7) admits a
solution
 $$\begin{aligned}
&\omega=-3\beta\alpha^2+\frac{(\vartheta_2^{(4)}\vartheta_1-\vartheta_1^{(4)}\vartheta_2)\alpha^3}
{\vartheta_1''\vartheta_2-\vartheta_2''\vartheta_1}\in\Lambda_0,\ \
\
c_1=\frac{(\vartheta_2^{(4)}\vartheta_1''-\vartheta_1^{(4)}\vartheta_2'')\alpha^3\sigma}
{\vartheta_1''\vartheta_2-\vartheta_2''\vartheta_1}\in\Lambda_1,
\end{aligned}\eqno(3.9)$$
here we have taken $f_1=\beta\sigma, \ \beta\in R$ for simplicity,
and other parameters $\alpha, \tau, \sigma$, $\beta$ are free.

By using (3.9) and solving
  system (3.8), we obtain that
 $$\begin{aligned}
&f_2=-\beta\omega\in\Lambda_0,\ \ \
c_2=\frac{(\vartheta_1^{(4)}\vartheta_2''-\vartheta_2^{(4)}\vartheta_1'')\alpha^3\omega}
{\vartheta_1''\vartheta_2-\vartheta_2''\vartheta_1}\in\Lambda_0.
\end{aligned}\eqno(3.10)$$

Noting that $\int\theta\ d\theta=1$ and $ \ \int d\theta=0$, we have
$$\partial^{-1} F_0=\int(\beta\sigma-\beta\omega\theta)\ d\theta=-\beta\omega.$$
 In this way,  we  indeed can get an explicit
periodic wave solution of the equation (1.12)
$$F=-\beta\omega+\partial_x\ln \vartheta(\xi,\tau),\eqno(3.11)$$
with the theta function  $\vartheta(\xi,\tau) $ given by (3.1) and
parameters $\omega$, $c_1,\ c_2$ by (3.9) and (3.10), while  other
parameters $\alpha, \sigma, \tau, \delta, \ \beta$  are free. Among
them, the three parameters $\alpha, \sigma$ and $\tau$ completely
dominate a periodic wave. In summary, the periodic wave (3.11) is
real-valued and bounded for all complex variables $(x, t, \theta)$.
 It  is one-dimensional, i.e. there is a single phase variable
$\xi$, and  has two fundamental periods $1$ and $i\tau$  in phase
variable $\xi$.

In the following, we further consider asymptotic properties of the
periodic wave solution. Interestingly,  the relation between the
one-periodic wave solution (3.11) and the one-super soliton solution
(1.5) can be   established  as follows.

{\bf Theorem 1.}  Suppose that the   $\omega\in\Lambda_0$ and
$c\in\Lambda_1$ are given given by (3.5), (3.9) and (3.10). For the
one-periodic wave solution (3.11), we let
$$\alpha=\frac{k}{2\pi i}, \ \ \sigma=\frac{\zeta}{2\pi
i},\ \ \delta=\frac{\gamma+\pi \tau}{2\pi i},\eqno(3.12)$$ where the
$k, \zeta$ and $\gamma$ are the same as those  in (1.5). Then we
have the following asymptotic properties
$$c\longrightarrow 0, \ \  \xi\longrightarrow\frac{\eta+\pi\tau}{2\pi i}, \ \
\vartheta(\xi,\tau)\longrightarrow 1+e^{\eta}, \ \ {\rm as } \ \
\lambda\rightarrow 0.\eqno(3.13)$$   In other words,  the  periodic
solution (3.11)  tends to the one-soliton solution (1.5)
 under a small amplitude limit , that is,
$$F\longrightarrow F_1, \ \ {\rm as } \ \
\lambda\rightarrow 0.\eqno(3.14)$$

{\it Proof.}  Here we will directly use the system (3.4) to analyze
asymptotic properties of one-periodic solution, which is more simple
and effective than  our original  method by solving the system
\cite{Dai}-\cite{Fan2} . Since the coefficients of system (3.4) are
power series about $\lambda$, its solution $(\omega, c)^T$ also
should be a series about $\lambda$.

  We
explicitly expand the coefficients of system (3.4) as follows
$$\begin{aligned}
&\vartheta_1(0,\lambda)=1+2\lambda^4+\cdots,\quad
\vartheta_1''(0,\lambda)=-32\pi^2\lambda^{4}+\cdots,\\
&\vartheta_1^{(4)}(0,\lambda)=512\pi^4\lambda^4+\cdots,
\ \ \vartheta_2(0,\lambda)=2+2\lambda^8+\cdots\\
&\vartheta_2''(0,\lambda)=-8\pi^2-72\pi^2\lambda^8+\cdots,\
\vartheta_2^{(4)}(0,\lambda)
=32\pi^4+2592\pi^4\lambda^8+\cdots.\end{aligned}\eqno(3.15)$$ Let
the solution of the system (3.4) be in the form
$$\begin{aligned}
&\omega=\omega_0+\omega_1\lambda+\omega_2\lambda^2+\cdots=\omega_0+o(\lambda),\\
&c=b_0+b_1\lambda+b_2\lambda^2+\cdots=b_0+o(\lambda),
\end{aligned}\eqno(3.16)$$
where $\omega_j\in\Lambda_0,\ b_j\in\Lambda_1, \ j=0,1,2\cdots$

Substituting the expansions (3.11) and (3.12) into the system (3.5)
and letting $\lambda\longrightarrow 0$, we immediately obtain
 the following relations
$$
 \begin{aligned}
 &b_0=0, \ \ -8\pi^2\sigma\omega_0
 +2b_0+32\pi^4\sigma\alpha^3=0,
 \end{aligned}$$
which has a solution
 $$b_0=0, \ \ w_0=4\pi^2\alpha^3.$$
Then from the relations (3.12) and (3.16), we have
$$c\longrightarrow 0, \ \  2\pi i\omega\longrightarrow 8\pi^3i\alpha^3=-k^3, \ \ {\rm as } \ \
\lambda\rightarrow 0,$$ and thus
$$\begin{aligned}
&\hat{\xi}=2\pi i\xi-\pi \tau=k x+2\pi i\omega t+\theta\zeta+\gamma\\
&\quad \longrightarrow kx-k^3t+\theta\zeta+\gamma=\eta,\ \ {\rm as}\
\ \lambda\rightarrow 0,
\end{aligned}\eqno(3.17)$$

 It remains to  show  that the one-periodic wave  (3.11) possesses  the same
form with the one-soliton solution (1.5) under the limit
$\lambda\rightarrow 0$. For this purpose, we first expand the
Riemann theta function $\vartheta(\xi, \tau)$ in the form
$$ \vartheta(\xi,\tau)=1+\lambda^2(e^{2\pi i\xi}+e^{-2\pi i\xi})+\lambda^8(e^{4\pi i\xi}+e^{-4\pi i\xi})
+\cdots .$$  By using the  (3.12) and (3.17),  it follows that
$$\begin{aligned}
&\vartheta(\xi,\tau)=1+e^{\hat{\xi}}+\lambda^4(e^{-\hat{\xi}}+e^{2\hat{\xi}})+\lambda^{12}(e^{-2\hat{\xi}}+e^{3\hat{\xi}})
+\cdots\\
&\quad \longrightarrow 1+e^{\hat{\xi}}\longrightarrow 1+e^{\eta},\ \
{\rm as}\ \ \lambda\rightarrow 0,
\end{aligned}$$
 which implies (3.13) and (3.14). Therefore we conclude that the one-periodic solution
(3.11) just goes to the one-soliton solution  (1.5) as the amplitude
$\lambda\rightarrow 0$. $\square$
\section{ Discussion on the conditions of  $N$-periodic wave solutions}

 In this section,  we consider condition for $N$-periodic wave solutions of
the equation (1.2).  The theta function is  taken  the form
 $$\vartheta(\boldsymbol{\xi}, \boldsymbol{\tau})=\vartheta(\xi_1, \cdots,\xi_N, \tau)
 =\sum_{\boldsymbol{n}\in \mathbb{Z}^N} \exp\{2\pi i<\boldsymbol{\xi},\boldsymbol{n}>-\pi
<\boldsymbol{\tau} \boldsymbol{n}, \boldsymbol{n}>\},\eqno(4.1)$$
where $\boldsymbol{n}=(n_1,\cdots, n_N)^T\in \mathbb{Z}^N,\ \
\boldsymbol{\xi}=(\xi_1, \cdots, \xi_N)^T\in \mathcal{C}^N,\ \
\xi_i=\alpha_jx+\omega_j t+\theta\sigma_j+\delta_j, \ \ j=1,\cdots,
N$, $\tau$ is a ${N\times N}$  symmetric positive definite matrix.

To  make  the theta function (4.1) satisfy the bilinear equation
(2.2), we  obtain that according to the formula (2.11)

 $$\begin{aligned}
  &\sum_{\mu=0,1}\ \ \sum_{n_1, \cdots, n_N=-\infty}^{\infty}
  G\left\{4\pi i\sum_{j=1}^{N}(n_j-\mu_j/2)\alpha_j, 4\pi
  i\sum_{j=1}^{N}(n_j-\mu_j/2)\omega_j,\right.\\
 &  \left.  4\pi i\sum_{j=1}^{N}(n_j-\mu_j/2)(\sigma_j+\theta\omega_j) \right\}\times \exp\left[-2\pi\sum_{j,
  k=1}^{N}(n_j-\mu_j/2)\tau_{jk}(n_k-\mu_k/2)\right]=0.
\end{aligned}\eqno(4.2)$$

Now we consider the number of equation and some unknown parameters.
Obviously, in the case of supersymmetric equations, the number of
constraint equations of the type (4.2) is $2^{N+1}$, which is  two
times of the constraint equations needed in the case of ordinary
equations \cite{Dai}-\cite{Fan2} . On the other hand we have
parameters $\tau_{ij}=\tau_{ji}, c_1, c_2,  f_1, f_2, \alpha_i,
\omega_i$, whose total number is $\frac{1}{2}N(N+1)+2N+4$. Among
them, $2N$ parameters $\tau_{ii}, \omega_i$ are taken to be the
given parameters related to the amplitudes and wave numbers (or
frequencies) of $N$-periodic waves; $\frac{1}{2}N(N+1)$ parameters
$\tau_{ij}$ implicitly appear in series form, which is general can
not to be solved explicit. Hence, the number of the explicit unknown
parameters is only $N+4$. The number of equations is larger than the
unknown parameters in the case when $N\geq 2$. In this paper, we
consider one-periodic wave solution of the equation (1.2), which
belongs to the cases when $N=1$.  There are still  certain
difficulties in the calculation for the case $N\geq 2$, which will be considered in our future work.\\[12pt]

\section*{Acknowledgment}

  I would like to express my special
thanks to the referee for constructive suggestions which have been
followed in the present improved version of the paper.  The work
described in this paper was supported by grants from the Research
Grants Council of Hong Kong
   (No.9041473), the National Science Foundation of China (No.10971031),
Shanghai Shuguang Tracking Project (No.08GG01) and Innovation
Program of Shanghai Municipal Education Commission (No.10ZZ131).

\end{document}